# Analysis of Diurnal Air Temperature Trends and Pattern Similarities in Highland and Lowland Stations of Italy and UK

Chalachew Muluken Liyew[1,4], Rosa Meo[1], Stefano Ferraris[2], Elvira Di Nardo[3*]

[1]Computer Science, University of Torino, Turin, Italy.
[2]Interuniversity Department of Regional and Urban Studies and Planning, Politecnico and University of Torino, Turin, Italy.
[3*]Mathematics "Giuseppe Peano", University of Torino, Turin, Italy. [4]Faculty of Computing, Bahir Dar University - Institute of Technology, Bahir Dar, Ethiopia.

*Corresponding author(s). E-mail(s): elvira.dinardo@unito.it;
Contributing authors: chalachewsweet@gmail.com; rosa.meo@unito.it; stefano.ferraris@unito.it;

## Abstract

In this paper, an analysis of hourly air temperatures in four groups of 32 stations of the UK highland (five stations), UK lowland (four stations), Italian highland (eleven stations), and Italian lowland (twelve stations) at various altitudes was conducted over the period from 2002 to 2021. The study aimed to examine the trends of each hour of the day in that period, over different averaging time windows (10-day, 30-day, and 60-day). The trends were computed using the Mann-Kendall trend test and Sen's slope estimator. The similarity of trends within and across the groups of stations was assessed using the hierarchical clustering with dynamic time warping technique. An additional analysis was conducted to show the correlation of trends among the group of stations using the correlation distance matrix. Hierarchical clustering and distance correlation analysis show trend similarities and correlations, also indicating dissimilarities among different groups. Using 30-day averages, significant warming trends in specific months at the Italian stations are evident, especially in February, July, August, and December. The UK highland stations did not show statistically significant trends, but clear pattern similarities were found within the groups, especially in certain months. The ultimate goal of this paper is to provide insights into temperature dynamics and climate change characteristics on regional and diurnal scales.

## 1. INTRODUCTION

Meteorological variables, collected extensively through hourly observations, are archived at many meteorological stations in different countries worldwide. This provides an opportunity for researchers to conduct useful research on climate change and its positive and negative effects on Earth's life. Some meteorological variables such as sensible heat flux, which is crucial for hydrology and climate studies [1], are expensive and complicated to measure directly [2]. They can be computed using the most available meteorological variables such as air temperature [2]. Numerous studies have also shown that precipitation is negatively or positively correlated with air temperature. The study in [3] revealed that less evaporative cooling and more sunshine are evident in dry conditions, while wet summers are cool. The authors of [4] investigated the relationships between surface air temperature and atmospheric, oceanic, and land surface variables showing the strong negative effect of altitude and humidity on air temperature. Therefore, the analysis of air temperature patterns and trends is a way to gain insights into other meteorological variables that are not measured easily.

For this study, observations recorded on an hourly basis in the UK and Italy from 2002 to 2021 are used to study mean temperatures in four groups of stations, at different altitudes. These groups include twelve stations (five in Valle d'Aosta and seven in Piemonte) in Italian lowlands, eleven stations (five in Piemonte and six in Valle d'Aosta) in Italian highland stations, five highland stations in the UK, and four low-altitude stations in the UK.

In analyzing temperature, a basic research focus is trend detection. For example, using linear regression, skewness, and variance methods, the diurnal and seasonal cycles of trends of air temperature were studied in [5]. The annual and seasonal trends were identified and quantified using the ordinary least square method in [6]. Using a multiple linear regression method, a long-term warming trend was found in [4] where the dependence on temperature time series on elevation was analyzed. A classical and robust method to verify the presence of a nonzero mean temperature trend and its statistical significance is the Mann-Kendall (MK) test (see e.g. [7–9]), also in the presence of seasonality [10]. On the other hand, Sen's slope estimator [7–9] is used to measure the magnitude of these trends. Compared with linear regression, MK, and Sen's slope estimator methods are less sensitive to extreme values and outliers and do not assume any particular distribution for the data. In any case, it is always advisable to check the population's normal distribution before applying these methods. Indeed, a single outlier may have a substantial impact on the estimation, potentially invalidating the trend interpretation [11]. In the following, the MK test is employed to detect the hourly temperature trend. The trend itself is quantified using Sen's slope estimator.

The analysis of air temperature trends was conducted on an hourly time scale. However, this scale has a too high resolution if patterns in trends aim to compare temperatures among different stations. Therefore, these data were grouped in different time windows such as 10-day, 30-day, and 60-day mean temperature. In literature, different studies aggregate air temperature observations on various time scales to analyze the trends or fluctuations in air temperature over specific time frames. Analyzing ERA5 reanalysis data from 1960 to 2021, the study in [12] verified that the hottest summer days in North-West Europe showed a warming trend about twice as fast as that observed on the average summer days. Monthly and annual mean air temperature trends were explored in [10] using daily mean air temperature data from both urban and coastal environments in Rome (Italy) and exhibited a statistically significant warming trend. Four-time scales such as monthly climate trends, at the beginning of the growing season, at the end of the growing season, and in the entire growing season were investigated in [7] using 302 observing stations in the US Midwestern states and homogenized historical monthly temperature records from 1980 to 2013. The result showed an increasing trend in the average temperature of the growing season in the regions with a different warming rate in the four-time scales. Seasonal and annual changes in maximum, minimum, and mean temperatures were assessed in [6] using a subset of 19 observatories from 1920 to 2003 and showed a statistically significant warming trend in maximum, minimum, and mean temperatures on both annual and seasonal scales. In [5] diurnal and seasonal cycles of surface air temperature trends and their variability were studied using hourly observations from nine geographically distributed stations in the US for the period 1951 - 1999. The diurnal asymmetry of surface air temperature was also studied using 45-year hourly observation data from Taiwan [13]. The presence of abrupt changes in time series is a very important topic, along with the study of their causes [14, 15]. However, we preferred to consider a short period for reasons of homogeneity of the sensors used in the measurements and to avoid nonlinearities in trends [16].

The ultimate goal of this study is primarily to examine the patterns and trends of the hourly temperatures within the Italian and the UK highland and lowland station groups in different time windows (10-day, 30-day, and two-way combinations of a 60-day mean). A further aim is to compare air temperature in the areas and altitudes represented by the four groups of stations. Thus hierarchical clustering was carried out using the dynamic time-warping (DTW) technique on diurnal mean temperature slopes within and among the four groups at the 30-day scale. The idea is to consider patterns that fall into the same cluster as similar. The distance correlation between the stations in a group is also explored to quantify the strength of relationships of diurnal temperature patterns at the 30-day scale.

The significance of this paper lies mainly in quantifying the existing signal in hourly temperature data with reference to the temporal variation over the past 20 years. This relatively limited time period is necessary as previously said, to avoid dishomogeneities and nonlinearity of trends in the Alpine region [16]. The analysis aims to detect the non-randomness of different trends along the year and along the day. As a matter of fact, a rapid increase in air temperature at ground level has been already recorded. But as the IPCC also remarks, this rapid increase is not homogeneous among different areas of the world. Thus, two very different areas of Europe were considered, one where the increase in temperature is relatively lower (northern United Kingdom) and one where it is relatively higher (northwestern Italian Alps). This difference is clearly shown in NOAA global data, but its patterns deserve to be studied in detail. Elevation effects [17] were also considered, analyzing different sets of stations at different altitudes, from 200 to 2000 m asl. As a matter of fact, the issue of altitude-dependent warming has been extensively studied, but quantifying its effects in relation to different geographic areas, either at a shorter distance (Piemonte vs. Valle d'Aosta) or a longer distance (Italy vs. the United Kingdom), provides useful insights.

The paper is organized as follows: the dataset is presented in Section 2 along with a summary of the methods employed for the analysis. The results and discussion are given in Section 3. The last section contains some concluding remarks.

## 2. MATERIAL AND METHODS
### 2.1. Study area and dataset

This study uses air temperature data collected from a total of 32 stations strategically placed in various geographical regions. These stations include both high and lowland areas in the UK and Italy. The dataset includes observations from five highland and four lowland stations in the UK, as well as eleven highland stations (five stations in Piemonte and six stations in Valle d'Aosta) and twelve lowland stations (five stations in Valle d'Aosta and seven stations in Piemonte) in Italy. The aim is to investigate and interpret the patterns observed in these different geographical locations through the analysis of trends in diurnal mean air temperature. The geographical areas of the study are visually represented in Figure 1.

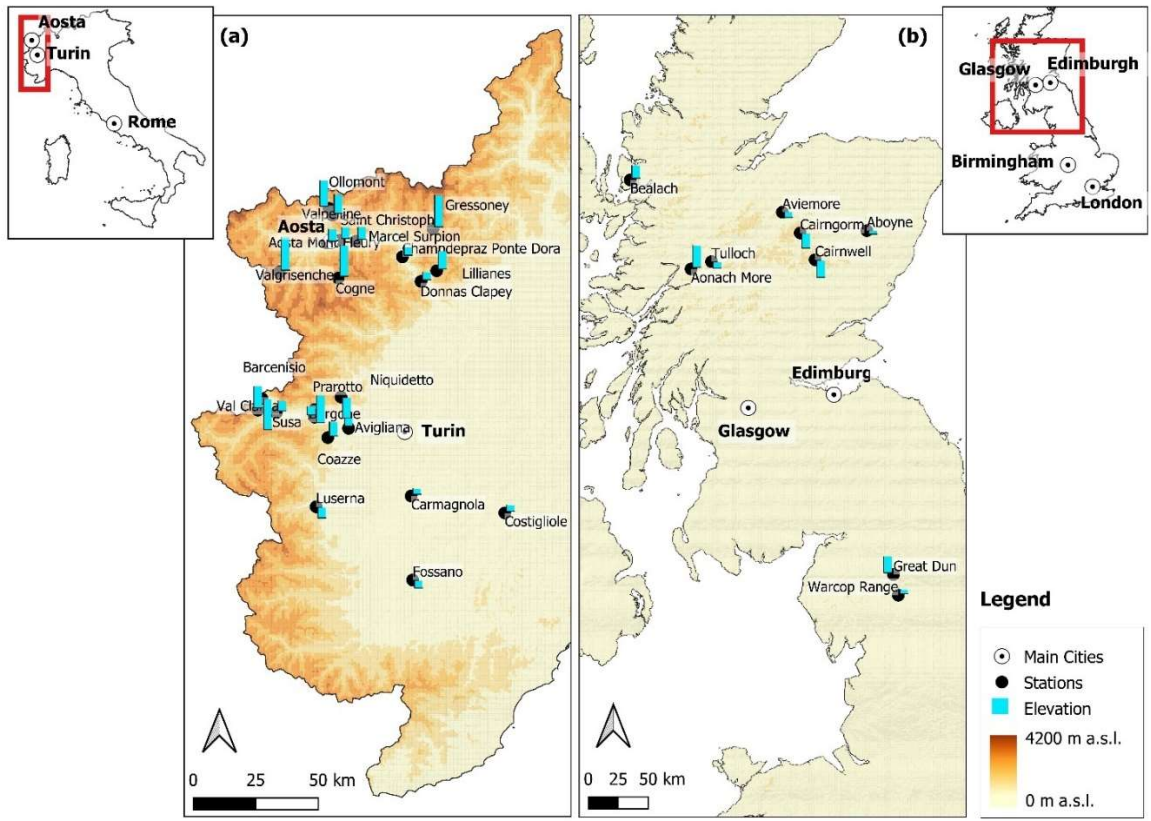

Fig. 1: The geographical distribution of all stations considered in this study: in (a) Italian highland and lowland, and (b) UK highland and lowland.

This study examines the patterns and trends of hourly mean air temperatures over 10-day, 30-day, and 60-day periods in the highland and lowland regions of the UK and Italy. By analyzing temperature variations at different temporal scales, this research aims to provide a comprehensive understanding of how climatic conditions vary across these regions. Comparing both highland and lowland areas will provide insights into how altitude can influence temperature dynamics and highlight potential differences in air temperature between the UK and Italy. The location and altitudes of each station of the dataset are given in Table 1.

Internal and natural variability is a huge problem in weather and climate modeling. In this paper, we aim to quantify the variation across regions and along the year and the day. However, we have to use individual stations that are not regularly and systematically distributed in space. Some studies considered the uneven distribution of the stations [18, 19, 20]. In addition, rapid and often nonlinear trends in the Alpine region [16] limit the survey to 20 years of data. Therefore, there are some limitations in this study, both spatial and temporal, and all efforts are directed toward quantifying the non-randomness of spatial and temporal variations despite these limitations.

Temperature records were collected at half-hourly intervals at all lowland Piemonte stations in Italy, totaling 350, 640 records. At high-altitude stations in Italy, the Valle d'Aosta lowlands in Italy, and the UK, as well as at all lowland stations in the UK, records were collected at one-hour intervals, for a total of 175, 320 records covering the period from 2002 to 2021. The effect of the missing values on the analysis of hourly mean air temperature is minimal since the dataset is sufficiently large and only a small number of missing values randomly appear. Special attention should be paid only to stations in the UK highlands that showed relatively consistent missing values. The last column of Table 1 shows the percentage value of missing data for each station.

An imputation technique was anyway used to get the most out of the dataset's readily accessible information, despite the relatively low frequency of missing values. In particular, the classic seasonally segmented missing-value imputation technique[1] was employed. When used as a pre-processing step, this method entails segmenting the time series into seasonal blocks, after which imputation is carried out using interpolation algorithms on an individual basis for each block. The dataset was thus transformed into an hourly mean time series after the imputed values were added. Such a strategy ensures that the imputed values maintain the inherent seasonal structure of the data.

---

[1] The function R na_seasplit() of the package imputeTS (version 0.3) was used to restore the missing information [21].

| Group | Region | Station location | Latitude | Longitude | Altitude (in m a.s.l.) | % Missing value |
|---|---|---|---|---|---|---|
| UK highland | | Cairngorm (CR) | 57.0607 °N | −3.6066 °E | 1237 | 10.94 |
| | | Aonach Mor (AN) | 56.8168 °N | −4.9603 °E | 1130 | 23.33 |
| | | Cairnwell (CW) | 56.8793 °N | −3.4213 °E | 928 | 13.60 |
| | | GreatDun (GD) | 54.6833 °N | −2.4500 °E | 847 | 7.74 |
| | | Bealach (BL) | 57.4167 °N | −5.7167 °E | 773 | 3.88 |
| UK lowland | | Aviemore (AR) | 57.2005 °N | −3.8282 °E | 228 | 0.58 |
| | | Aboyne (AY) | 57.0767 °N | −2.7803 °E | 140 | 1.46 |
| | | Tulloch (TH) | 56.8667 °N | −4.7067 °E | 249 | 0.29 |
| | | WarcopRange (WR) | 54.5344 °N | −2.3900 °E | 227 | 0.85 |
| Italian lowland | Valle d'Aosta | Saint Christophe (SC) | 45.7393 °N | 7.3634 °E | 545 | 0.58 |
| | | Champdepraz Ponte (CP) | 45.6818 °N | 7.6737 °E | 370 | 1.70 |
| | | Marcel Surpion (MS) | 45.7366 °N | 7.4446 °E | 540 | 0.75 |
| | | Donnas Clapey (DC) | 45.5966 °N | 7.7664 °E | 318 | 0.58 |
| | | Aosta Mont Fleury (MF) | 45.7305 °N | 7.2990 °E | 577 | 0.32 |
| | Piemonte | Luserna (LS) | 44.8084 °N | 7.24601 °E | 475 | 0.08 |
| | | Susa (SS) | 45.1386 °N | 7.0484 °E | 470 | 0.09 |
| | | Costigliole (CT) | 44.7866 °N | 8.1822 °E | 440 | 0.03 |
| | | Fossano (FS) | 44.5496 °N | 7.7251 °E | 403 | 0.06 |
| | | Borgone (BG) | 45.1229 °N | 7.2380 °E | 400 | 0.22 |
| | | Avigliana (AG) | 45.0841 °N | 7.4071 °E | 340 | 0.12 |
| | | Carmagnola (CM) | 44.8462 °N | 7.7177 °E | 232 | 0.16 |
| Italian highland | Piemonte | Valclarea (VC) | 45.1477 °N | 6.9567 °E | 1068 | 0.11 |
| | | Prarotto (PR) | 45.1490 °N | 7.2370 °E | 1431 | 0.14 |
| | | Niquidetto (NI) | 45.1937 °N | 7.3692 °E | 1416 | 0.10 |
| | | Coazze (CO) | 45.0515 °N | 7.3039 °E | 1130 | 0.15 |
| | | Barcenisio (BA) | 45.1880 °N | 6.9774 °E | 1525 | 0.79 |
| | Valle d'Aosta | Gressoney (GR) | 45.7796 °N | 7.8258 °E | 1642 | 0.29 |
| | | Cogne (CG) | 45.6083 °N | 7.3561 °E | 1682 | 0.48 |
| | | Valgrisenche (VG) | 45.6297 °N | 7.0640 °E | 1859 | 0.24 |
| | | Ollomont (OL) | 45.8494 °N | 7.3102 °E | 2017 | 0.42 |
| | | Lillianes (LL) | 45.6337 °N | 7.8442 °E | 1256 | 0.13 |
| | | Valpelline (VP) | 45.8263 °N | 7.3273 °E | 1029 | 0.34 |

**Table 1**: Location, latitude, longitude, altitude (in m a.s.l.), and missing value percentage of the weather stations where the air temperature was registered.

## 2.2. Methods

In this subsection, we shall briefly review the nonparametric methods employed for identifying and measuring hourly temperature trends within various time frames, (including 10-day, 30-day, and a combination of two 60-day means) and for existing pattern similarity measures among the trends.

***Non-parametric trend analysis***. The MK test is a widely employed non-parametric method for analyzing the statistical significance of a trend in time series [22], commonly used in fields like hydrology and climatology [8, 23]. The MK test[2] finds a trend in a time series without identifying whether the trend is linear or nonlinear. The null hypothesis of no trend is tested against the alternative hypothesis that the trend is positive or negative. In particular, the MK test evaluates how the signs of earlier and later data points differ. The idea is that sign values tend to either constantly increase or decrease if there is a trend. Each data point in the time series is compared to every other value, resulting in a total of N (N − 1)/2 pairs of comparisons, if N is the sample size. If the p-value test is less than 0.05, there is statistically significant evidence of a trend in the time series. Before running the test, the data were examined for serial correlation to avoid biased conclusions and confirmed that there was no serial correlation in the diurnal mean temperature, as the chance of rejecting the null hypothesis even if there is no trend [25]. The magnitude of the trend is measured by Sen's slope[3] estimator [8, 9]. The method is robust to outliers and might be appropriately used for nonnormally distributed censored time series with missing values. Sen's slope estimator is calculated as the median of all slopes $T_i = (X_i - X_j)/(j - k)$ for $i = 1,2, \ldots, N$. Positive values of this med.an indicate an upward or increasing trend whereas negative values indicate a downward or decreasing trend.

***Hierarchical clustering with Dynamic Time Warping.*** The Dynamic Time Warping (DTW) procedure is an unsupervised machine learning method used to measure the similarity between two-time series $X = \{x_1, x_2, \ldots, x_n\}$ and $Y = \{y_1, y_2, \ldots, y_m\}$, also with $n \neq m$, by finding the optimal alignment (warp path) between X and Y, involving a one-to-many mapping for each pair of elements. To do this, DTW searches for the optimal alignment that minimizes the distance between corresponding points [26].

The details of the procedure can be found in Algorithm 1. In the following, we briefly recall the relevant steps. The first step consists in computing a matrix $C$, where $C(i,j)$ is the cost of the pair $(x_i, y_j)$, that is the cost of aligning two points in the series by considering various possible alignments through dynamic programming. Indeed, the following "chess king moves" are possible: $(i,j) \rightarrow (i, j+1)$ (horizontal), $(i,j) \rightarrow (i+1, j)$ (vertical), and $(i,j) \rightarrow (i+1, j+1)$ (diagonal). The matrix is initialized by setting $C_{0,0} = 0$ and $C_{i,0} = \infty, C_{0,j} = \infty$. The values

---

[2] The R function mk.test() of the package trend (version 1.1.5) was used [24]
[3] The R function sans. slope() of the package trend was used for the san's slope method

$C(i,j)$ are then calculated recursively by using a suitable distance function $d(x_i, y_j)$, and taking into account the following constraints on the warping paths (see the second for the cycle in Algorithm 1):

  a) the alignment starts at pair (1, 1) and ends at pair (N, M);
  b) the order of the elements in $X$ and $Y$'s path should be maintained;
  c) a pair $(x_i, y_j)$ can be followed by the three possible pairs $(x_{i+1}, y_j)$, $(x_i, y_{j+1})$ and $(x_{i+1}, y_{j+1})$.

More in details we have

$$C(i,j) = d(x_i, y_j) + \min \begin{cases} C(i-1, j-1) & \text{diagonal move} \\ C(i-1, j) & \text{horizontal move} \\ C(i, j-1) & \text{vertical move} \end{cases}$$

where $d(x_i, y_j)$ is an appropriate distance function. After computing $C$, the optimal alignment can be found by backtracking from $C_{n,m}$ to $C_{0,0}$. To manage the alignments, the matrix $DTW$ is computed from $C$ where $DTW(i,j)$ is the distance between two sub-sequences $\{x_1, \ldots, x_i\}$ and $\{y_1, \ldots, y_j\}$. The values $DTW(i,j)$ are calculated recursively according to the three possible moves:

$$DTW(i,j) = \min \begin{cases} DTW(i-1, j-1) + wd \times C(i,j) & \text{diagonal move} \\ DTW(i-1, j) + wh \times C(i,j) & \text{horizontal move} \\ DTW(i, j-1) + wv \times C(i,j) & \text{vertical move} \end{cases}$$

where *wh, wv,* and *wd* are the weights corresponding to the horizontal, vertical, and diagonal moves respectively. When all weights of horizontal, vertical, and diagonal moves are equal to one, the recursive function favors diagonal alignment since the cost of a single diagonal step is lower than the combined cost of one vertical and one horizontal step. As suggested by the literature, choosing (*wh, wv, wd*) = (1, 1, 2) reduces this bias. The final DTW distance is the total cost of the best warp path and indicates how well the two sequences can be aligned at the lowest possible cost. Smaller DTW distances indicate greater similarity between the sequences, as they require less distortion to align optimally. DTW is prone to overfitting, a problem that arises when, for example, the warping window is not appropriately selected in equal-length sequences. This can lead to inflated similarity scores between sequences. To overcome this issue, a regularization technique has been employed by incorporating a penalty term into the cost function. This penalty term aims to penalize large warping steps. The result is a corrected DTW cost functions

$$DTW_{regularized}(i,j) = DTW(i,j) + \lambda \times \gamma(i,j)$$

where $\gamma(i,j)$ denotes the regularization term whose strength is tuned by the parameter $\lambda$. With this choice, alignment steps that have a large difference in indices are penalized, discouraging the alignment from jumping too far off the diagonal. Taking the average of the distances, the hierarchical clustering is then applied. The following strategy is of agglomerative bottom-up type.

This method groups similar data points according to their features. At the beginning, each data point is assigned to its own cluster. Then the procedure iteratively merges the closest clusters until a desired number of clusters is achieved or all data points are merged into a single cluster.

---

Algorithm 1: Algorithm to measure the similarity of two time series

**Input:** $X$ and $Y$
**Cost Matrix:** $C \in R^{(n+1)\times(m+1)}$
**Initialization:**

  for $i = 1$ to $n$: $C_{i,0} = \infty$
  for $j = 1$ to $m$: $C_{0,j} = \infty$
  $C_{0,0} = 0$

**Calculate the cost matrix:**

for $i = 1$ to $n$ do
  for $j = 1$ to $m$ do

$$C(i,j) = d(x_i, y_j) + \min \begin{cases} C(i-1, j-1) & \text{diagonal move} \\ C(i-1, j) & \text{horizontal move} \\ C(i, j-1) & \text{vertical move} \end{cases}$$

  end for
end for

**Get Alignment:** Trackback from $C_{n,m}$ to $C_{0,0}$

---

***Silhouette*** Score. To evaluate the performance of hierarchical clustering, the Silhouette Score[4] has been computed [27]. This score measures the similarity of a data point in its own cluster to other clusters by comparing the mean intra-cluster distance with the mean nearest-cluster distance. More in detail the Silhouette Score $S(i)$ for the data $x_i$ is defined as

$$S(i) = \begin{cases} \dfrac{b(i) - a(i)}{\max(a(i), b(i))} & \text{if } |C_I| > 1 \\ 0 & \text{if } C_I = 1 \end{cases}$$

where $C_I$ denotes a cluster, $|C_I|$ its size and

$$a(i) = \frac{1}{|C_I| - 1} \sum_{j \in C_I, i \neq j} d(i,j) \qquad b(i) = \min_{J \neq I} \left\{ \frac{1}{|C_J|} \sum_{j \in J} d(i,j) \right\}$$

---

[4] The R function *Silhouette()* was used from the package clusters.

with $d(i,j)$ the distance between $x_i$ and $x_j$ within the same cluster. Note that $a(i)$ is a measure of the mean intra-cluster distance of $x_i$ to all other data belonging to $C_I$. Instead, $b(i)$ is the minimum mean distance of $x_i$ to another cluster $C_J$ with $J \neq I$, and so identifies the closest neighboring cluster of $x_i$. The overall Silhouette Score S is the mean of all $S(i)$ values and ranges from −1 to 1. A value close to 1 indicates well clustered data points and a value around 0 indicates data points on or near cluster boundaries. A negative value is likely to indicate that the data points would be better distributed in a nearby cluster rather than in their current cluster.

Distance correlation. Distance correlation[5] is a statistical measure used to assess relationships between two random variables or sets of variables [29]. It does not require the normality assumption, in contrast to the Pearson correlation, and it is invariant to both linear and some non-linear transformations. The distance covariance (dcov) and distance correlation (dcor) are derived from Euclidean distances between sample pairs. The process involves transforming sample distances and normalizing squared distances. The result is a versatile measure for capturing dependencies in data beyond linear relationships ranging from 0 (no correlation) to 1 (perfect correlation).

## 3. RESULTS AND DISCUSSION

Air temperature is aggregated for each hour of the day over a 10-, 30-, and 60-day period to analyze diurnal temperature patterns, understand temperature fluctuations, and identify trends over time. It is particularly useful for assessing diurnal variations in temperature and for understanding how temperatures change at different times of day in a group of stations. The primary objective of this section is to examine the similarity of long-term temperature patterns and slopes (trends) at different altitudes within the Italian and UK highlands and lowlands.

The time series patterns of the hourly mean air temperature from 2002 to 2021 for each group of stations are shown in Figure 2. Hourly mean air temperature ranges from $0°C$ to $5°C$ in the UK highlands, from $5°C$ to $12°C$ in the UK lowlands, from $0°C$ to $15°C$ in the Italian highlands and from $5°C$ to $19°C$ in the Italian lowlands. The hourly mean temperature patterns in the lowlands are similar, however, the Donnas and Champdepraz stations show different slope structures from their reference groups, as do Valpelline and Lillianes show higher hourly mean temperatures with different slopes than their reference groups. A well-known fact is further observed, namely that the hourly average temperature at all stations is highest around noon and lowest in the morning and evening.

In the four groups of stations, the hourly mean air temperature has a minimum value at 5 hours and peaks between 12 and 14 hours. Both lowland and highland stations in the UK show similar patterns within their groups. Conversely, among stations in, the Italian highland, Valpelline, and Lillianes have higher hourly mean temperatures than their counterparts. Hourly mean temperature patterns at lowland stations in Italy are consistent among the stations.

---

[5] The R function *dcor.test()* of the package energy (version 1.7 – 11) was used [28]

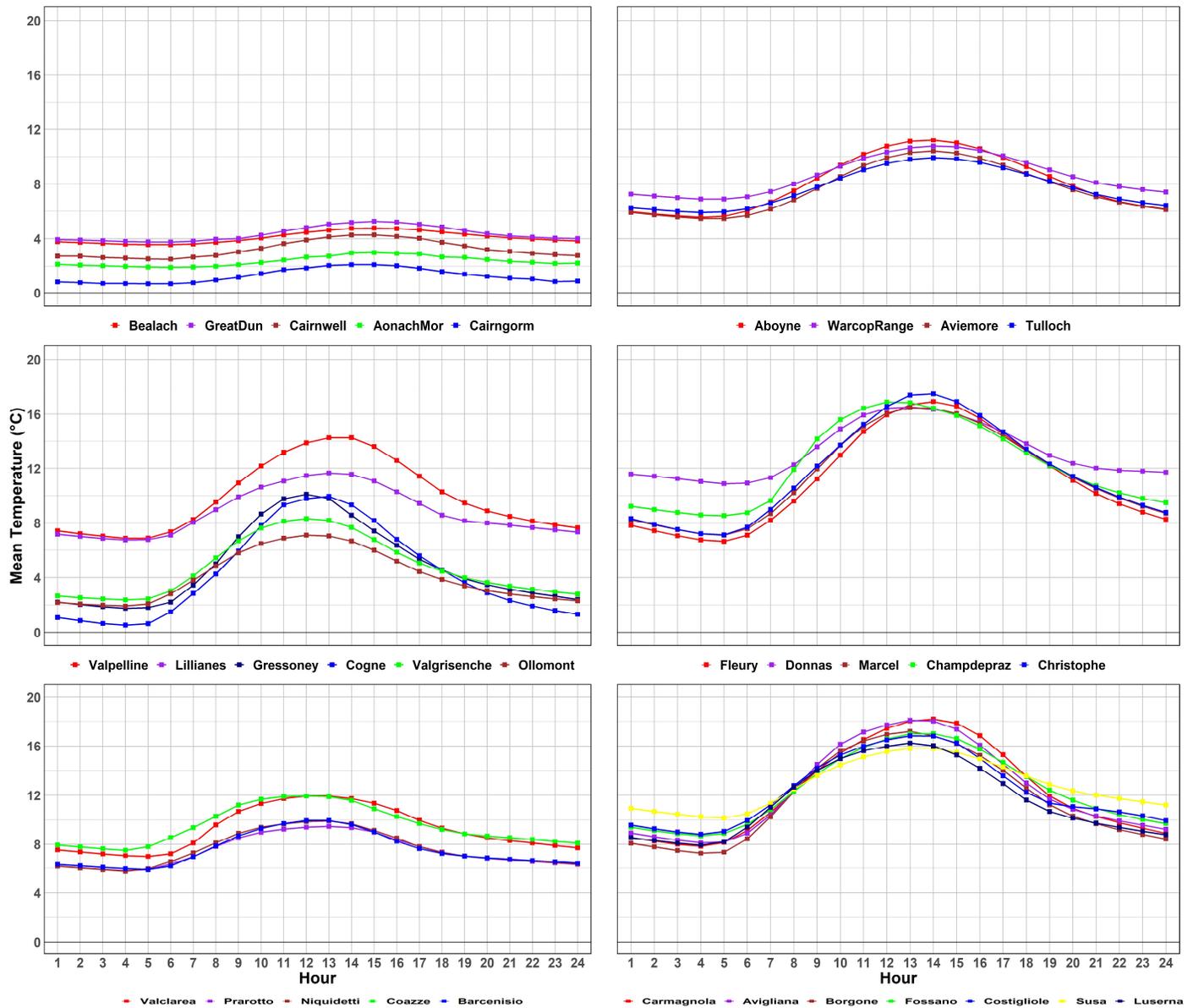

**Fig. 2**: The time series patterns of hourly mean temperature from 2002 to 2021 in UK stations (first row plots) and Italian ones (The Valle d'Aosta station plots are in the second row and Piemonte station plots are in the third row.)

Further insights into the meteorological processes have been obtained from the analysis of hourly mean temperature trends, which has been initially performed at 30 days. Therefore, the trend of hourly mean temperatures is computed in each month at every hour. The slope of diurnal temperatures at each group of stations in each month is computed for the period from 2002 to 2021. Figure 3 shows the monthly plots for June and December. Plots of the other 10 months are shown in the supplementary material (see Fig. S.1). The slopes show similarity within each group and each month, although some stations show slightly different patterns.

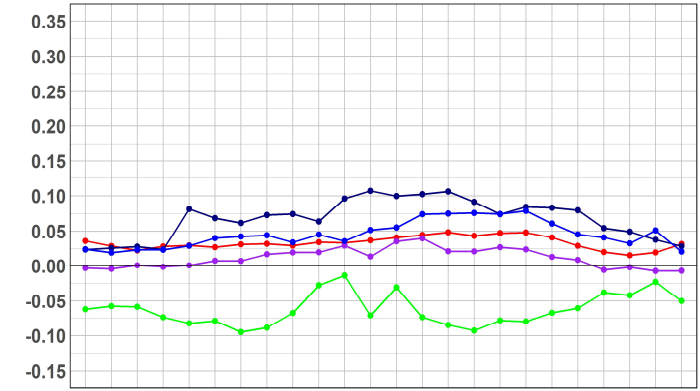
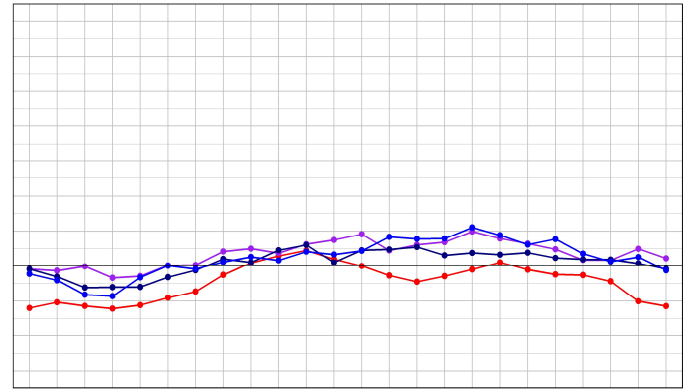
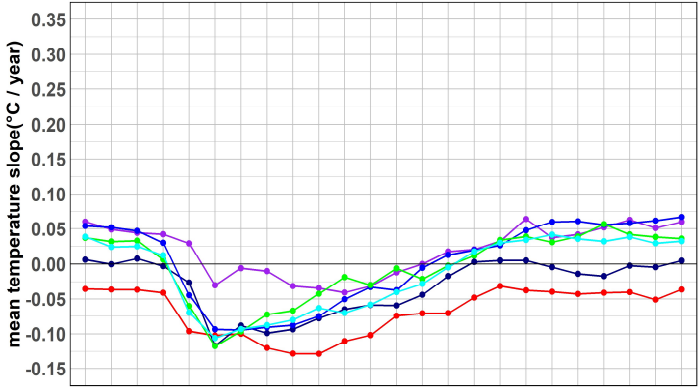
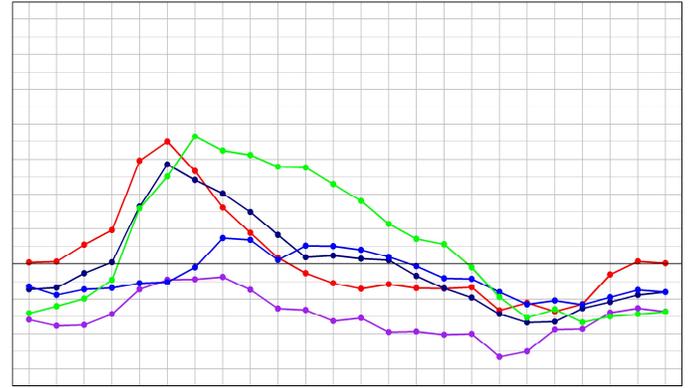
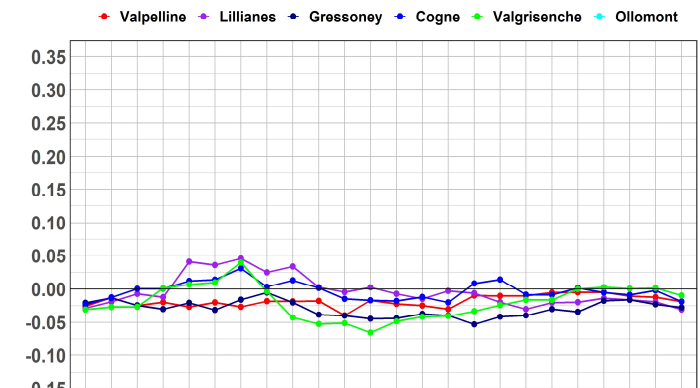
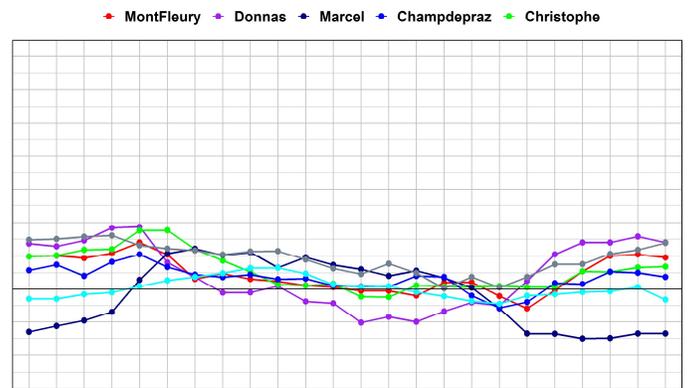

(a) June

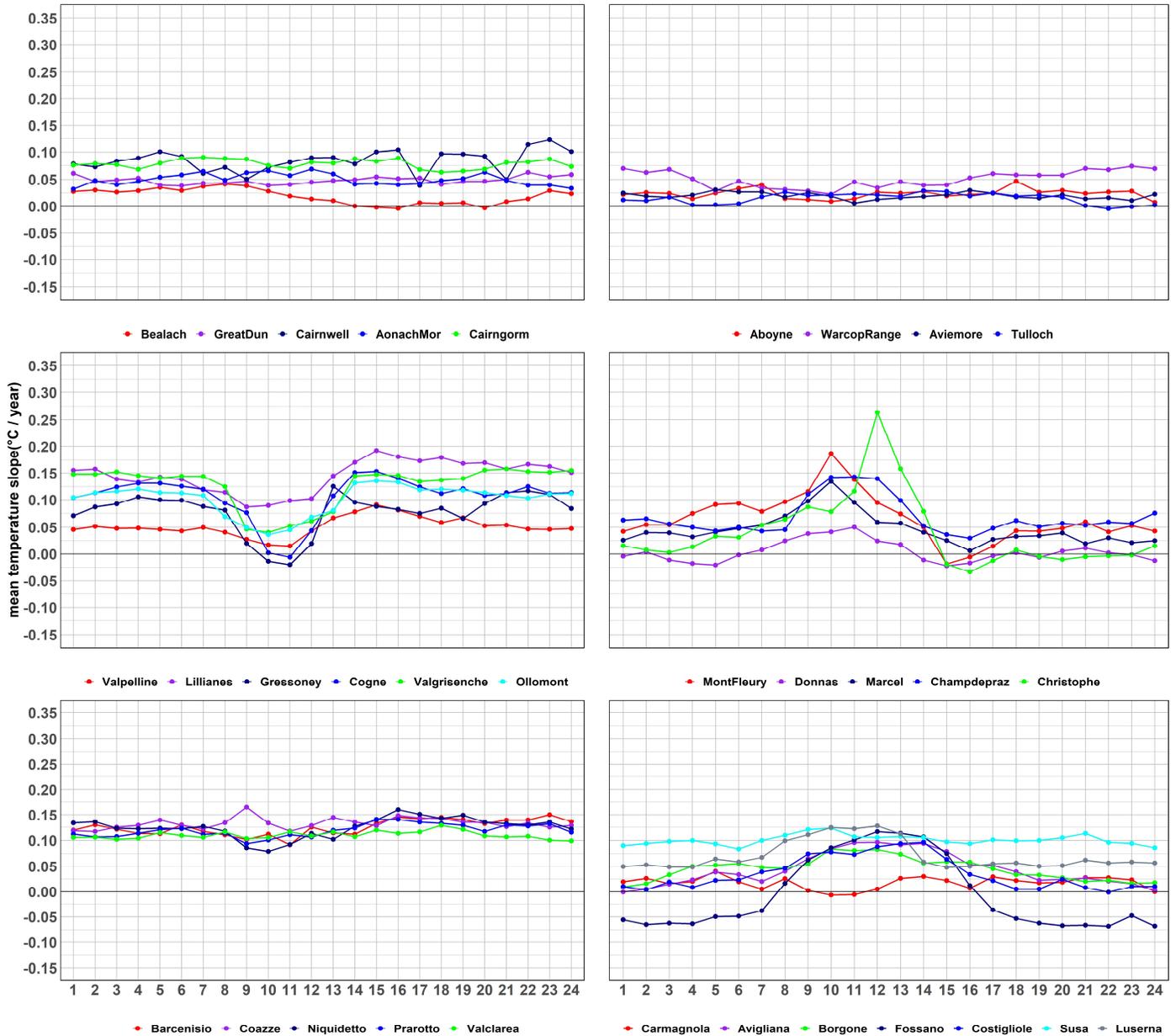

(b) December

**Fig. 3**: Slopes of the hourly mean temperature trends in June (a) and December (b) from 2002 to 2021. The first row contains the UK stations, the second row contains the Valle d'Aosta stations, and the last row contains the Piemonte stations.

The similarity of diurnal mean temperature slopes (see Table 2) within and among the groups are investigated using hierarchical clustering in conjunction with the DTW method on hourly temperatures. Note that this analysis is new in the literature of this field. Four clusters have been chosen, taking into account the geographic areas involved in the data set and the shape of the dendrograms built with the hierarchical clustering. Thus, for each month, stations of the four geographic areas are assigned to one of the four clusters.

| Month | Group | Cluster I | Cluster II | Cluster III | Cluster IV | S. Score |
|---|---|---|---|---|---|---|
| January | IH | VP, OL, VG, GR, NI, BA, PR, VC, CO | LL, CG | | | 0.54 |
| | IL | DC, SC | CM, AG, BG, CT, LS, CP, MF, MS, SS | FS | | |
| | **UKH** | **AN** | | | **CR, CW, BL, GD** | |
| | UKL | | | | WR, TH, AY, AR | |
| February | IH | VP | LL, OL, CG, PR, GR, VC, BA, CO, VG, NI | | | 0.46 |
| | IL | CM, BG, CT, SS | AG, LS, DC | FS | SC, CP, MS, MF | |
| | **UKH** | **AN** | | **CR, BL, GD, CW** | | |
| | UKL | | | WR, TH, AY, AR | | |
| March | IH | VP | LL, OL, VG, CG, GR, CO, VC, BA, NI, PR | | | 0.43 |
| | IL | CT, SS, DC | CM, AG, BG, LS | FS | MS, MF, CP, SC | |
| | **UKH** | **BL, GD, CW, CR** | **AN** | | | |
| | UKL | WR, TH, AY, AR | | | | |
| April | IH | VP, OL, CG, VG, GR, NI, BA, VC | LL, PR, CO | | | 0.44 |
| | IL | CM, BG, CT, SS, DC, MF, CP | AG, LS | FS, MS, SC | | |
| | **UKH** | | | | **BL, GD, CW, CR, AN** | |
| | **UKL** | | | | WR, TH, AY, AR | |
| May | IH | VP, NI | LL, OL, CG, VG, GR, BA, VC, PR, CO | | | 0.29 |
| | IL | CM, CT, DC, FS | AG, BG, LS, SS, MF | MS, SC, CP | | |
| | **UKH** | | **BL, CW, CR, GD** | | **AN** | |
| | UKL | | WR, TH, AY, AR | | | |
| June | IH | VP, LL, OL, CG, VG, GR | | BA, NI, VC, PR, CO | | 0.32 |
| | IL | | CM, CT, AG, BG, LS | FS, SS, MF, MS, SC, CP | DC | |
| | **UKH** | | **AN, BL, CW** | **GD** | **CR** | |
| | UKL | | | WR, TH, AY, AR | | |
| July | IH | VP, GR, BA, VC, PR, CO, NI | LL OL, CG, VG | | | 0.31 |
| | IL | CM, CT, AG, BG, LS, MF, FS, SS, CP | | MS, SC, DC | | |
| | **UKH** | **AN, BL** | | **CW, GD** | **CR** | |
| | UKL | | | WR, TH, AY, AR | | |
| August | IH | VP, GR, NI, VC | LL OL, CG, VG, BA, PR, CO | | | 0.54 |
| | IL | CM, CT, AG, BG, LS, MF, FS, SS, CP, DC | | MS, SC | | |
| | **UKH** | | | | **BL, GD, CW, CR, AN** | |
| | UKL | | | | WR, TH, AY, AR | |
| September | IH | VP, OL, CG, GR | LL, VG, BA, NI, VC, PR, CO | | | 0.51 |
| | IL | DC, CP | CM, CT, AG, BG, LS, SS, | FS, MF, MS, SC | | |
| | **UKH** | | | | **BL, GD, CW, CR, AN** | |
| | UKL | | | | WR, TH, AY, AR | |
| October | IH | VP, CG, GR, NI, VC, CO | LL, OL, VG, BA, PR | | | 0.44 |
| | IL | CM, CT, AG, BG, LS, SS, CP, DC | | FS | MF, MS, SC | |
| | **UKH** | **AN, BL, CW, CR, GD** | | | | |
| | UKL | WR, TH, AY, AR | | | | |
| November | IH | VP | LL | OL, CO, VC, VG, CG, GR, BA, NI, PR | | 0.43 |
| | IL | | CM, CT, AG, BG, LS, SS, CP | | MF, MS, SC, FS, DC | |
| | **UKH** | **AN, BL** | | **CW** | **CR, GD** | |
| | UKL | WR, TH, AY, AR | | | | |
| December | IH | VP | LL, VG, BA, NI, PR, CO | OL, CG, GR, VC | | 0.32 |
| | IL | AG, CM, BG, LS, MF, MS, SC, CP, CT, DC | | SS | FS | |
| | **UKH** | **AN, BL, GD** | | | **CW, CR** | |
| | UKL | WR, TH, AY, AR | | | | |

**Table 2**: Classification of the 32 stations into four clusters using hierarchical clustering in conjunction with DTW for the IH, IL, UKH, and UKL stations. The values of the Silhouette Score are given in the last column.

The performance of this procedure is evaluated using the Silhouette Score, as shown in the last column of Table 2. The Silhouette Scores vary considerably from month to month, ranging from 0.54 in January and August to 0.29 in May. The most defined and consistent clusters are observed at the months with higher scores, while lower scores lead to more overlapping and less distinct clusters. One of the most prominent observations is the consistency of the clustering for the UK lowlands. In all months, UK lowland stations (AY, TH, WR, AR) are grouped in the same cluster (usually Cluster I and IV) most frequently, and Cluster III and Cluster II occasionally. This indicates a strong similarity in the characteristics of these stations, which remain unchanged throughout the year (see also Figure 2). The UK highlands group appears more heterogeneous. In April, August, September, and October, all UK high-land stations consistently exhibited similar clustering patterns, grouped predominantly in Cluster IV three times and in Cluster I once. Conversely, in January, February, March, and May, all UK highland stations cluster together in the same group, except for Aonach Mor, which is grouped differently. Italian lowland and highland stations show distinct and varied clustering behaviors throughout the year in comparison with UK ones. Stations grouped in the same clusters are dynamically changed in each month. However, they are most frequently grouped into two clusters (Cluster I and Cluster II). Valpelline appears consistently in Cluster I in all months, alone or together with a few stations of its reference group. This suggests that Valpelline has unique features that strongly differentiate it from other stations. Most of the stations in the Piemonte highlands group together consistently, while the Valle d'Aosta highlands also often cluster in the same group, with the exception of the Valpelline stations, as mentioned above. For the Italian lowland stations, no consistent clustering is observed within its group. Fossano shows a highly dynamic clustering pattern throughout the year, often appearing in a cluster distinct from the other stations that highlights its distinctive characteristics. Occasionally Fossano shares clusters with some stations of its group, particularly with the Piemonte stations such as Carmagnola, Costigliole, Avigliana, Borgone, Luserna, and Susa, and the Valle d'Aosta stations such as Mont Fleury and Champdepraz Ponte.

To simplify the reading of the clustering given in Table 2 and add some further insights, radar charts are shown in Figure 4 for each month. In each plot, lines (red, black, orange, and green) identify the clusters. The spokes represent the six groups, that is UK highlands and lowlands, Italian highlands and lowlands (Piemonte and Valle d'Aosta). Thus, compared with Table 2, we stratified the analysis on Italian stations between Piemonte and Valle d'Aosta. The length of a spoke is proportional to the corresponding Silhouette Score average. Confirming the interpretation of Table 2, there is great variability among stations, particularly for Cluster I, which mainly collects Italian stations. Cluster II exhibits less variability. UK lowland and UK highland stations exhibit similar behavior quite different from Italian stations. Piemonte and Valle d'Aosta highland stations are more similar, with a few exceptions such as June and December. Dissimilarity is most evident between Piemonte and Valle d'Aosta lowland stations. The months with the least variability are May, October and December. In these cases, some discrepancy from the geographical area of a few stations is recorded, in particular for Cluster II in May and Cluster I in October.

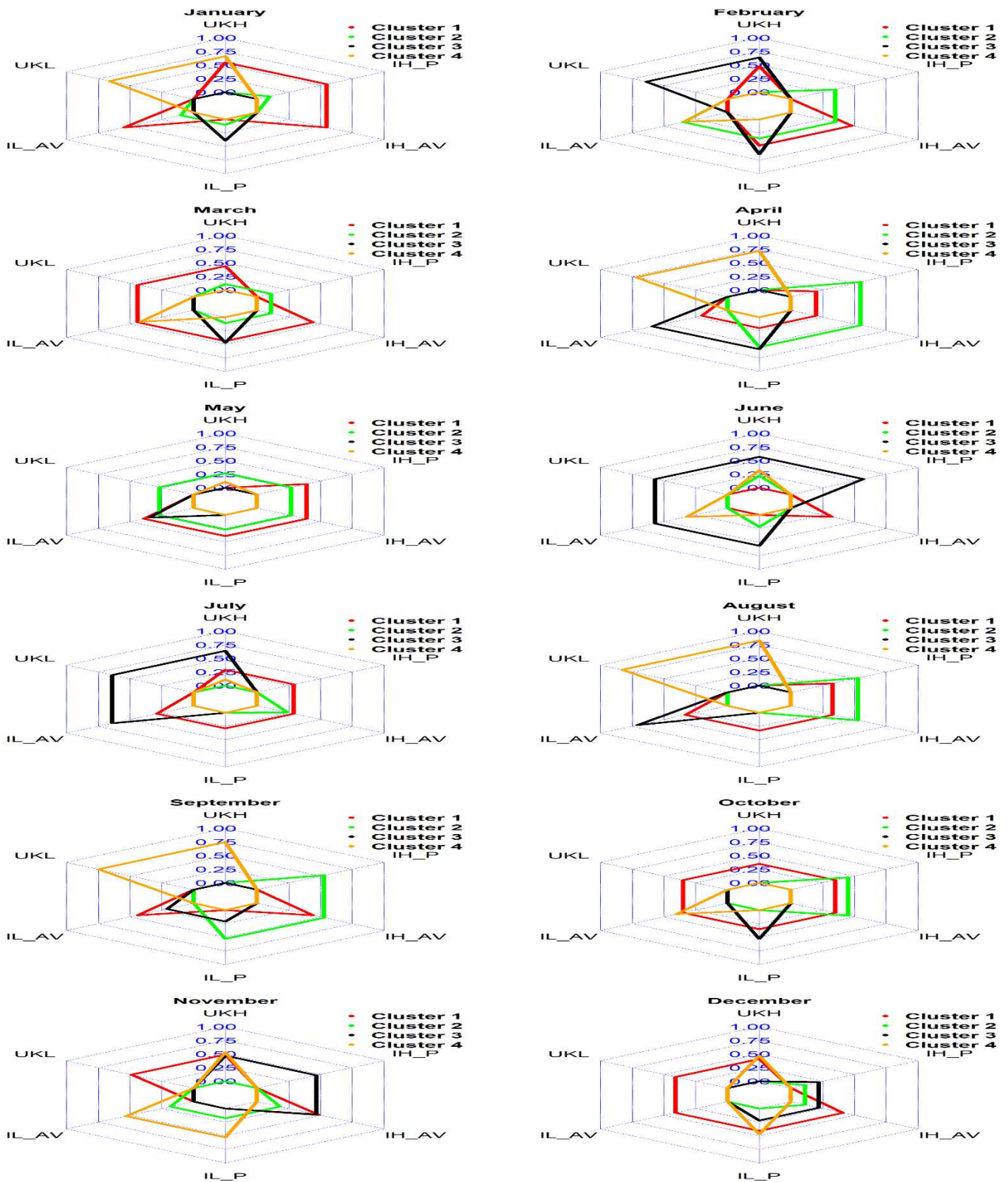

Fig. 4: Radar charts of the Silhouette Score averages for the six regions (UKH is UK Highland, UKL is UK Lowland, IL_AV is Italian Lowland Valle d'Aosta, IL_P is Italian Lowland Piemonte, IH_AV is Italian Highland Valle d'Aosta, IH_P is Italian Highland Piemonte) of the four geographical location groups. The red, green, black, and orange lines correspond to Cluster I, II, III, and IV respectively, as given

in Table 2.

The distance correlation matrix of the 32 stations in each month is depicted in Figure 5. Notably, the Italian highland stations showed a relatively high correlation with its group in all months. The UK highland stations are also highly correlated with its own group stations in some months, for example, in the interval between March and June. The Italian lowland station Fossano shows a special feature being correlated with its group as well as with most of the Italian highland stations in some. As shown in Figure 5, the stations show a relatively high correlation within their own group. However, stations with special characteristics, such as Fossano, Sussa and Aonach Mor, showed a relatively higher correlation with a few stations outside their own group. In June, the Italian lowlands have a relatively lower correlation with the UK highlands and lowlands, with the exception of Fossano and Susa which are correlated with the Italian highlands. In April, May, and June, there is a significant correlation among the UK lowlands. The same happens for the UK highlands whose correlation is meaningful also with the UK lowlands.

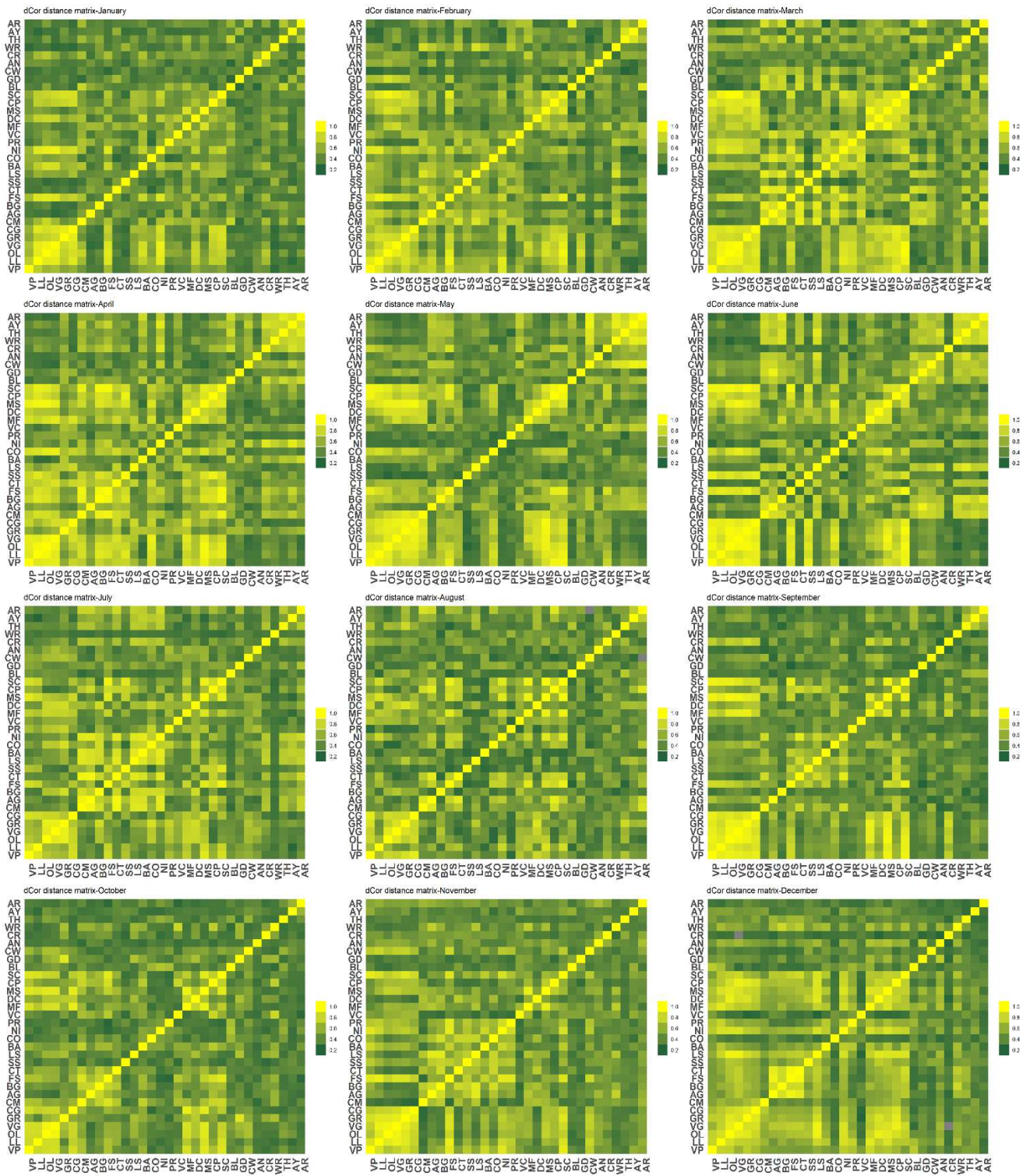

**Fig. 5**: The dCor distance matrix for the slopes of the hourly mean temperature of the 32 stations of Italy (highland and lowland) and the UK (highland and lowland).

### 3.1. Choosing a day scale

Trends observed in the hourly mean temperature time series are the subject of a further in-depth study. Indeed, the mean temperature is calculated at intervals of 10-, 30-, and 60- days for each hour of the day. To study the 10-day scale, the month is divided into three groups, each of which includes 10 days (1-10, 11-20, and 21-31 days). The mean temperature for each hour of the day is then calculated for these 10-day periods. Similarly, to explore the 30-day scale, the mean temperature is calculated for each hour of the day over 30 days. For the 60-day scale, two alternative combinations of months are considered:

- *i)* the first combination includes January-February, March-April, May-June, July-August, September-October and November-December,
- *ii)* the second combination consists of December-January, February-March, April-May, June-July, August-September, and October-November.

To carry out the analysis, all of these mean temperatures are subsequently converted into vectors and time series to which the methods described in Section 2.2 are applied.

To summarize the results carried out by running the MK test and Sen's slope estimator to these new datasets, Figures from 6 to 9 (as well Figures from S.2 to S.5 in the supplementary materials) show suitable contour plots of the hourly mean temperature slopes and *p*-values. The aim is to show the significant change in hourly temperature in a month. In particular, Figure S.2 refers to the 10-day scale, Figure S.3 to the 30-day scale, and Figure S.4 and Figure S.5 to the 60-day scale in the two combinations described above, namely *i)* and *ii)* respectively. The contour levels of the *p*-values (surface) expressed as a function of hours (*y-axis*) and months (*x-axis*) are displayed in the second column of each of these figures, whereas the slopes are presented in the first column. As the computed *p*-values vary within the interval $(0.001, 1]$, the partition of this interval into the three subintervals $(0.001, 0.05]$, $(0.05, 0.10]$, $(0.10, 1]$, is indicated by the different colors of the areas, making it easier to identify the critical ones. The second column of all these figures shows the contour levels obtained by expressing the slopes as a function of months (*x-axis*) and hours (*y-axis*). As the computed slopes vary in $(-1.0, 1.0]$, the partition of this interval into the four subintervals $(-1.0, -0.03]$, $(-0.03, 0.0]$, $(0.0, 0.03]$, $(0.03, 1.0]$ is again indicated by the different colors of the areas. As a result, the slopes that match the central intervals are the least significant.

These plots provide valuable insights into the temperature patterns of each station. Indeed, the following key points can be highlighted. Significant warming trends are observed in the late summer, fall, and winter at the highland and lowland stations in Italy. These trends vary significantly when examining particular hours and months in each group. In particular, the most significant months on the 30-day scale are July, August, and September; on the 10-day scale, the most significant months are February, August, and December. Cooling trends are instead observed in the spring. Different warming and cooling trends during particular combinations of months are also displayed at the 60-day scale. There is undoubtedly a general warming trend in February, July, August, and December. Unlike the Italian stations, those in the UK show more stable temperatures. In particular, the UK highlands show no statistically significant trend at the 10-day, 30-day, and 60-day scale. The exception is April for the UK lowland stations since all-day scale windows show a trend toward cooling. In the following, some more details are given on the analysis of the contour plots at the different scales.

**The 30-day scale.** The contour plots of Italian highlands show a statistically significant warming trend from June to October in most stations from the evening to the morning hours with a slope ranging in (0.03, 1] except Valpelline (see Figure [6](#)) which showed a cooling trend. For a few hours in the morning, Cogne and Gressoney experienced a noteworthy cooling trend that is comparable in magnitude. Moreover, from the evening to the morning of the same day in August, September, and October, the contour plots of the majority of the Italian highland stations show statistically significant warming trends, with a slope ranging in (0.03, 1]. In Italian lowlands, statistically significant warming trends are consistently observed from evening to late morning during the summer season. Also in November, a notable warming trend appears with different patterns at different stations. In some of the remaining months, a complex warming trend can be inferred at various times of the day. Note that the observed statistically significant warming trends show a slope ranging from 0.03 to 1.0. For the majority of the UK's highland stations, the contour plots show no statistically significant trend for each month (Figure S.3 (A) in the supplementary material). Bealach station is worth mentioning since a statistically significant cooling trend is visible in January between 12 and 24 (hours). The UK lowland stations exhibit a slightly different behavior. Indeed a statistically significant cooling trend is observed in April and January (see Figure S.3 in the supplementary material) with the exception of WarcopRenger.

**The 10-day scale.** According to Figure S.2 in the supplemental material, the contour plots show comparable statistically significant warming trends in December for both the Italian highland and lowland stations, occurring the whole day for most stations and a few hours of a day in some stations. The warming trend is also observable in the Italian highlands and lowlands in February and April excluding Susa. February is a tricky month for the Italian lowlands since a complex pattern of warming trends appears in the contour plots. However, at the Gressoney (Italian highland) stations and Fossano (Italian lowland), a statistically significant cooling trend is observed in April at particular times of the day. For the UK highlands, no significant trends are detectable at any of the stations. Instead, at four stations of UK lowlands, April corresponds to a statistically significant cooling trend in the morning.

**The 60-day scale** This day scale shows statistically significant warming trends for two distinct periods, as shown in Figures S.4 and S.5 in the supplementary material: the July-August combination and the August-September combination. These warming trends persist from the evening through the next morning. The only station where this pattern does not appear to be significant is Valpelline in the combination of August- September. The Italian highlands and lowlands exhibit a significant warming trend at all stations during the September–October period. Nonetheless, the trend changes at various hourly intervals. In the November-December combination, a warming trend is observed in three stations of the Italian highlands (Lillianes, Valgrisench, and Cogne). All Italian lowland stations exhibit the same behavior, except Carmagnola, which does not exhibit any trend at any time of day during this month's combination. Trends are noted in Bealach during the January- February period, in GreatDun during the March- April combination, and in Cairnwell during the same period. Aboyne discloses a cooling trend during the night and morning sessions for the March–April and May–June combination. Furthermore, Cairngorm stations exhibit a warming trend in November and December, especially during the day.

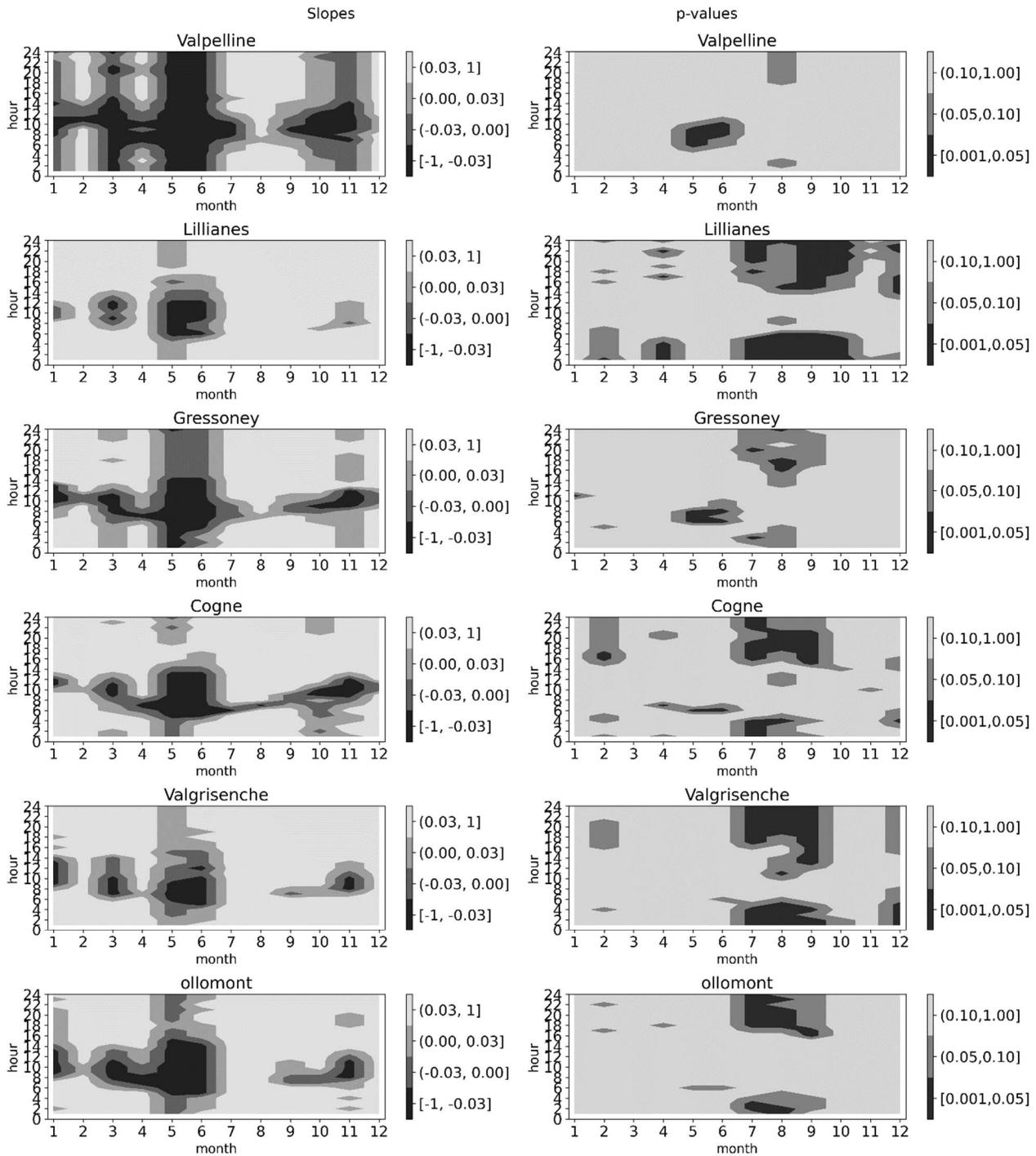

**Fig. 6**: The contour plots of the Valle d'Aosta highland mean temperature **at** the 30 - day with slopes in the first column and p-values in the second column.

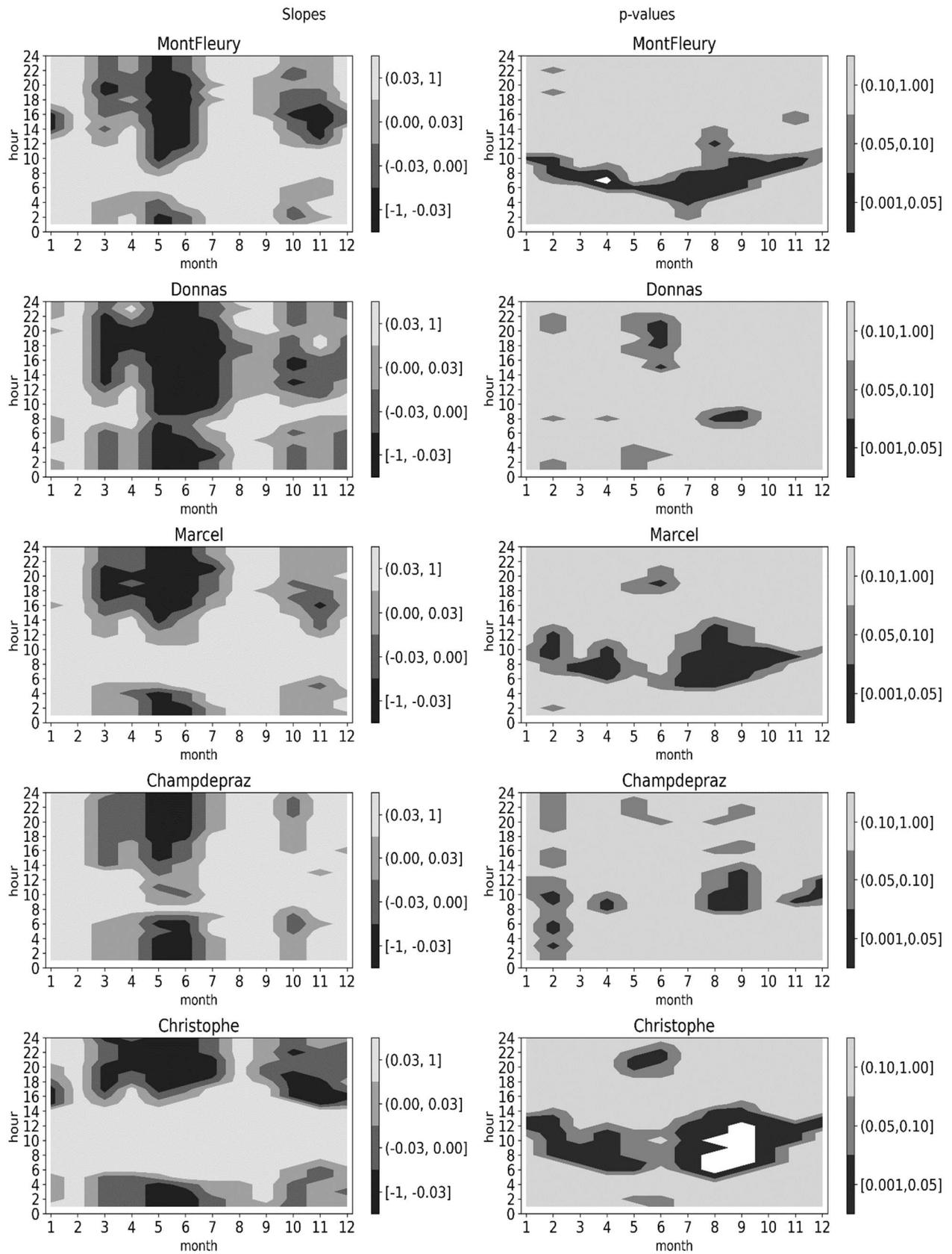

**Fig. 7**: The contour plots of the Valle d'Aosta lowland mean temperature **at** the 30- day with slopes in the first column and p-values in the second column.

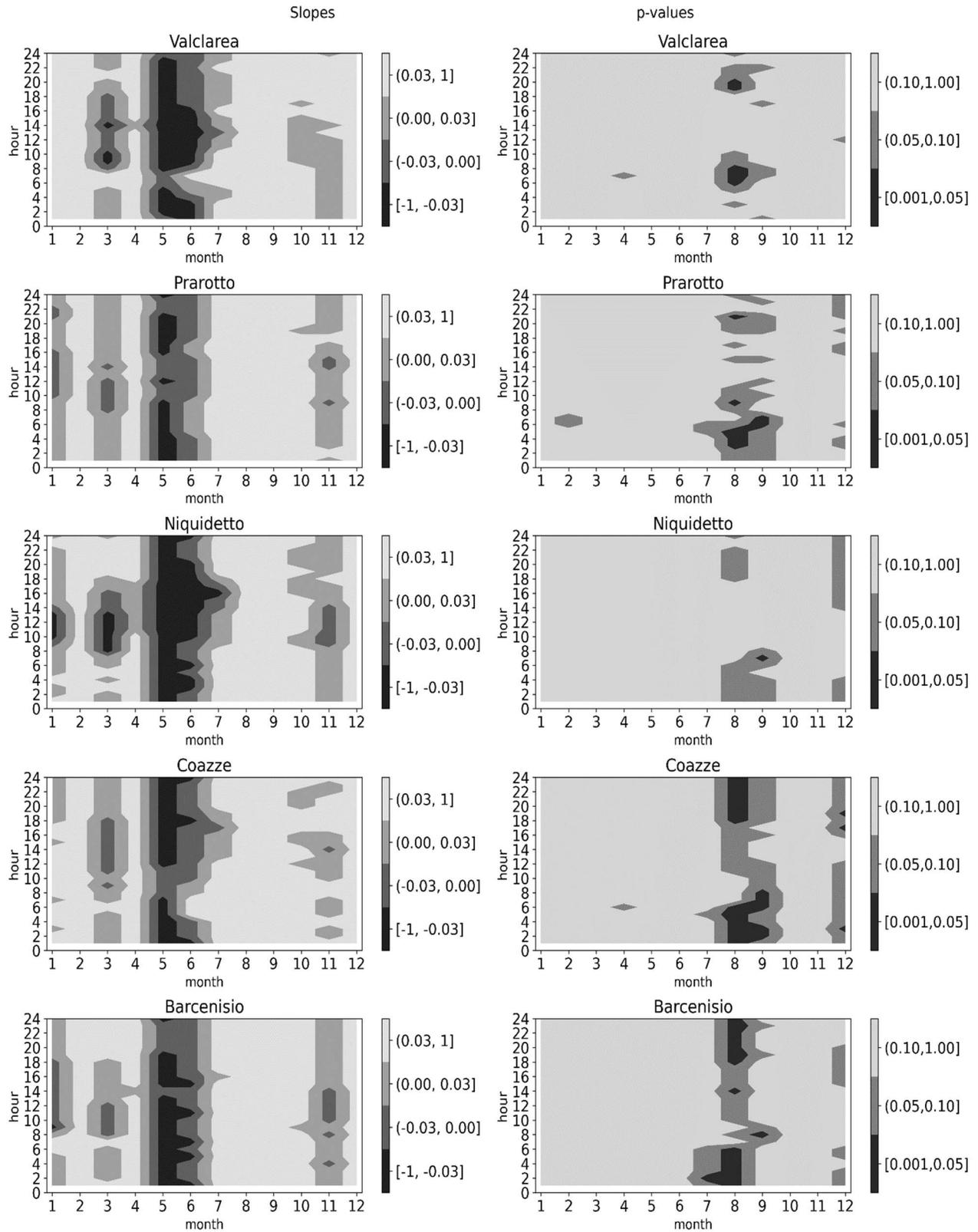

**Fig. 8**: The contour plots of the Piemonte highland mean temperature **at** the 30-day with slopes in the first column and p-values in the second column.

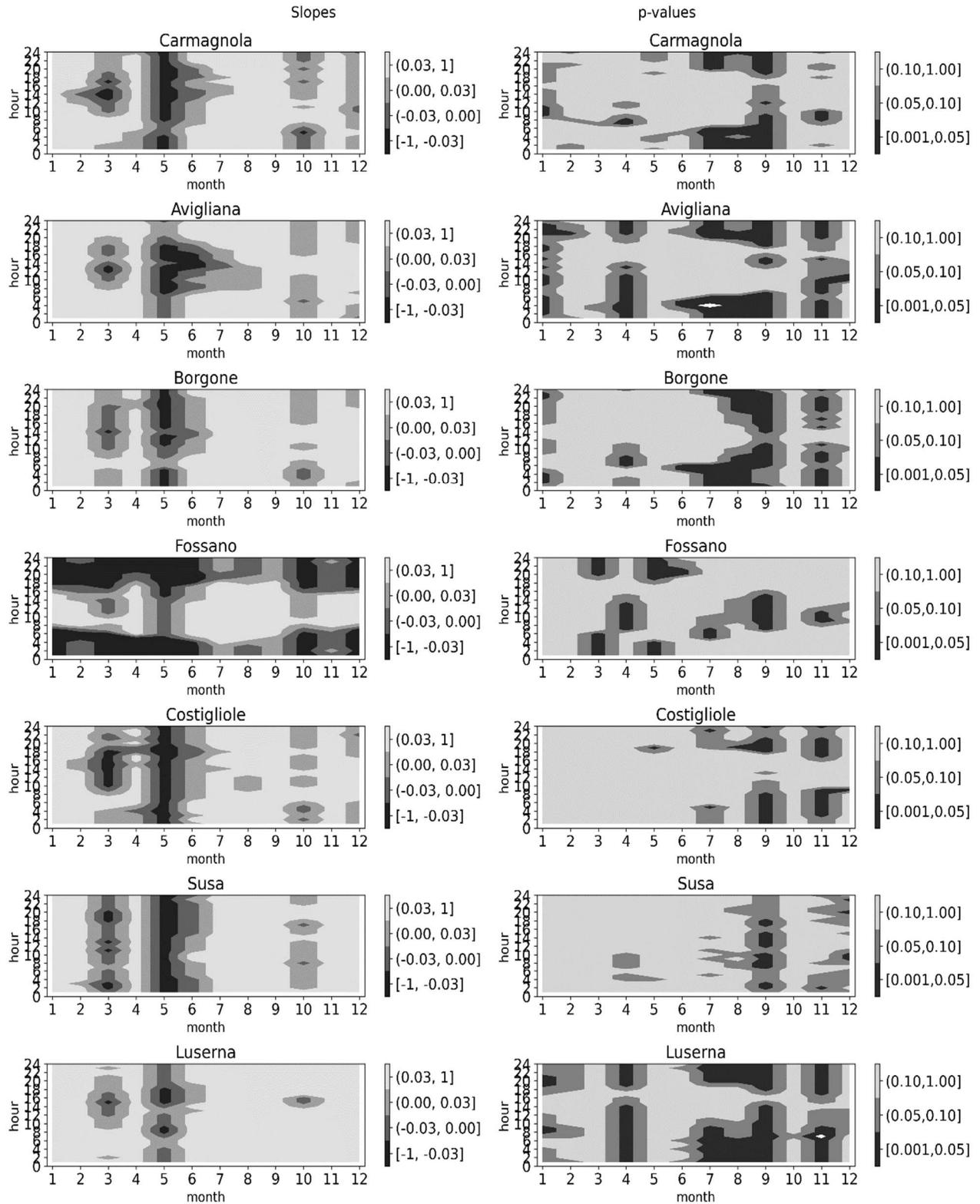

**Fig. 9**: The contour plots of the Piemonte lowland mean temperature **at** the 30-day with slopes in the first column and p-values in the second column.

To compare our results with those in [5] we repeated their analysis in the different cases of Italy and the UK. We found less randomness in Italy (see the contour plots in Figures 6 to 9) than in their graphs. This finding is interesting because our results refer to a much shorter time series (in [5] time series range from 1951 to 1999). These differences may be due to the stronger effects of climate change, especially in the Alps. In fact, in Figures S.2 to S.5 of the supplementary material, the UK contour plots are noisier.

4. CONCLUSION

In this paper, a comprehensive analysis of hourly air temperatures is presented considering 32 stations located at various altitudes and divided into four groups: UK highlands (five stations), UK lowlands (four stations), Italian highlands (six Valle d'Aosta and five Piemonte stations ), and Italian lowlands (seven Piemonte and five Valle d'Aosta stations). The time period considered covers the years from 2002 to 2021. The study aims to examine trends in each hour of the day, by averaging values over different time windows (10-day, 30-day, and 60-day).

The UK highlands and lowlands have shown cooling trends in most months of each hour of the day. The Italian highland and lowland stations have registered positive slopes in most months. However, there is variability in trends throughout the day. In some months this variability shows similarity in each group of stations.

Using DTW and the hierarchical clustering unsupervised machine learning technique, the hourly temperature pattern similarity of values and trends has been examined both within and between the groups. Most stations in the same groups are classified into the same cluster.

Distance correlation analysis shows that stations in the Italian highlands are relatively highly correlated with their group of stations in all months. There is also a high correlation with Fossano (Italian plain), with which there is also somewhat similarity.

Specific temperature trends are detected for each group of stations in Italian lowland and highland groups. There are distinct patterns, with significant warming trends observed in February and December. The duration of these patterns varied among the stations starting in the afternoon and lasting into the early morning hours. A similar warming trend can be seen in December at Italian highland stations. Conversely, no significant trend is observed in the UK highlands, but the stations show similar trend patterns to each other. This study provides insights into temperature patterns for each station, emphasizing the contribution of specific months (December, July, August, and February) to the overall warming trend for the group of Italian stations.

A combination of different processes can cause these diurnal patterns but to study these processes is beyond the scope of this paper. The findings of the present paper highlight the need to understand temperature dynamics in different regions and altitudes of Europe and emphasize the importance of continuous monitoring and analysis to assess the characteristics of climate change.

**Supplementary information.** For this study, a file is added as supplementary materials to show the comprehensive analysis here conducted, for different situations. The first figure (See Figure S.1.) is the scatter plot at the 30-day scale. The slopes of the hourly mean temperature at the same scale are depicted in each month except June and December which are shown in Figure 3 in the paper. The second issue in the supplementary material (see Figure S.2 - Figure S.5) presents 10-, 30- and 60-day mean temperature trends, depicted as contour plots with hourly data. The *p*-values in the second column and the slopes in the first column have been computed using the statistical methodology outlined in Section 2.2.

**Acknowledgements.** The authors would like to acknowledge the funder of this paper. This publication is part of the project NODES which has received funding from the MUR – M4C2 1.5 of PNRR funded by the European Union - NextGenerationEU (Grant agreement no. 470 ECS00000036). It was also partially funded by the PRIN 2022 project Snow Droughts Prediction in the Alps: A Changing Climate. We also would like to express our sincere gratitude to the reviewers whose insightful comments and constructive suggestions have significantly improved the quality of the paper.

Declarations

- Funding: MUR – M4C2 1.5 of PNRR funded by the European Union - NextGener- ationEU (Grant agreement no. 395 ECS00000036) and also partially funded by the PRIN 2022 project Snow Droughts Prediction in the Alps: a Changing Climate.
- Conflict of interest/Competing interests: We declare that we have no competing interests.
- Ethics approval and consent to participate: Not applicable
- Consent for publication: Not applicable
- Data availability: Temperature data are freely available from MetOffice, CFVDA, and ARPA Piemonte. The results of this paper are available upon request from the authors.
- Materials availability: Not applicable
- Code availability: The R code for the experiment is available in the footnotes of the paper
- Author contribution: Conceptualization: S.F, C.M.L., and E.D.N.; Design: S.F., C.M.L., E.D.N., and R.M.; Data curation: S.F. and C.M.L; software programs: C.M.L; Methodology: E.D.N. and C.M.L.; Analysis and Interpretation of results: S.F., C.M.L., E.D.N.; Draft manuscript preparation: C.M.L, E.D.N, and S.F.; Review and editing: E.D.N., R.M., and S.F.; Financial sourcing: S.F. All authors read and approved the final version of the manuscript.